\documentclass[12pt,paper=A4]{scrartcl}
\usepackage[utf8]{inputenc}
\usepackage[american]{babel}
\usepackage{mathtools} 
  \mathtoolsset{centercolon} 
\allowdisplaybreaks[3]
\usepackage{amsthm,amssymb}
\usepackage{expl3}
\usepackage{tikz} 
    \usetikzlibrary{calc,intersections}
    \tikzset{%
      myLine/.style={line width=#1*0.93pt},
      myLine/.default=1,
      myFace/.style={fill=gray!20}}
\definecolor{myColor}{HTML}{887733}
\usepackage{hyperref}
\usepackage[rightcaption]{sidecap}

\DeclareMathOperator{\dist}{dist}
\DeclareMathOperator{\tr}{tr}
\DeclareMathOperator{\range}{ran}

\DeclarePairedDelimiter{\abs}{\lvert}{\rvert}
\DeclarePairedDelimiter{\norm}{\lVert}{\rVert}
\DeclarePairedDelimiter{\tnorm}{\lVert}{\rVert}
\DeclarePairedDelimiter{\commutator}{\lbrack}{\rbrack}
\DeclarePairedDelimiter{\braket}{\langle}{\rangle}
\DeclarePairedDelimiter{\bra}{\langle}{\rvert}
\DeclarePairedDelimiter{\ket}{\lvert}{\rangle}
\DeclarePairedDelimiterXPP{\trace}[1]{\tr}{\lparen}{\rparen}{}{#1}

\DeclarePairedDelimiter{\paren}{\lparen}{\rparen}

\DeclarePairedDelimiter{\List}\{\}
\providecommand\given{}

\newcommand\SetSymbol[1][]{%
    \nonscript\,#1\ordinarycolon
    \allowbreak
    \nonscript\,
    \mathopen{}}
\DeclarePairedDelimiterX\Set[1]\{\}{%
    \renewcommand\given{%
        \SetSymbol[\delimsize]}
    \nonscript\,
    #1
    \nonscript\,
}

\def\ifempty#1{\def\temp{#1} \ifx\temp\empty}

\newcommand{\sumstack}[2][]{%
    \ifempty{#1}
        \sum_{\substack{#2}}
    \else
        \smashoperator[#1]{\sum_{\substack{#2}}}
    \fi
}

\providecommand{\todo}[2][To-Do]{{\color{olive}(\textbf{#1:} #2)}\PackageWarning{To-Do}{#1: #2}}
\renewcommand{\todo}[2][]{} 
\newcommand{\LTadd}[1]{} 
\newcommand{\LTskip}[1]{#1} 
\LTadd{\newcommand{\todo}[2][]{}} 

\renewcommand{\Phi}{\varPhi}

\renewcommand{\Lambda}{\varLambda}

\renewcommand{\Gamma}{\varGamma}

\renewcommand{\Omega}{\varOmega}

\newcommand{\E}{{\mathrm{e}}}

\newcommand{\N}{ \mathbb{N} }
\newcommand{\Z}{ \mathbb{Z} }
\newcommand{\HSpace}{\mathcal{H}}
\newcommand{\unit}{\mathbf{1}}

\newcommand{\Alg}{\mathcal{A}}
\newcommand{\domainH}{\mathcal{D}}
\newcommand{\Aloc}{\mathcal{A}_\mathrm{loc}}

\newcommand{\thmheadfont}[1]{{\normalfont\sffamily\bfseries#1}}
\newcommand{\thmcitefont}[1]{{\normalfont\sffamily#1}}

\newtheoremstyle{theorem} 
{}                    
{}                    
{\itshape}                    
{}                    
{\sffamily\bfseries}  
{.}                   
{ }                   
{}                    

\newtheoremstyle{definition} 
{}                    
{}                    
{}                    
{}                    
{\sffamily\bfseries}  
{.}                   
{ }                   
{}                    

\theoremstyle{theorem} 
\newtheorem{prop}{Proposition}
\newtheorem{thm}[prop]{Theorem}
\newtheorem{lem}[prop]{Lemma}
\newtheorem{cor}[prop]{Corollary}

\theoremstyle{definition}
\newtheorem{defi}[prop]{Definition}

\title{Local stability of ground states in  locally gapped and weakly interacting quantum spin systems}

\author{Joscha Henheik\footnote{Institute of Science and Technology Austria (IST Austria), Am Campus 1, 3400 Klosterneuburg, Austria. E-Mail: joscha.henheik@ist.ac.at}, \
    Stefan Teufel\footnote{Mathematisches Institut,
        Eberhard Karls Universit\"at T\"ubingen, Auf der Morgenstelle 10, 72076
        T\"ubingen, Germany. 
        E-Mail: stefan.teufel@uni-tuebingen.de}, \
    Tom Wessel\footnote{Mathematisches Institut,
        Eberhard Karls Universit\"at T\"ubingen, Auf der Morgenstelle 10, 72076
        T\"ubingen, Germany. 
        E-Mail: tom.wessel@uni-tuebingen.de}}

\begin{document}

\maketitle

\begin{abstract}
    Based on  a result by Yarotsky (J.~Stat.~Phys.~118, 2005), we prove that localized but otherwise arbitrary perturbations of weakly interacting quantum spin systems with uniformly  gapped on-site terms change the ground state of such a system only locally, even   if they close the spectral gap.
    We call this a \emph{strong version} of the \emph{local perturbations perturb locally} (LPPL) principle which is known to hold  for much more general gapped systems, but only for perturbations that do not close the spectral gap of the Hamiltonian.
    We also extend this strong LPPL-principle  to Hamiltonians that have the appropriate structure of gapped on-site terms  and weak interactions only locally in some region of space.

    While our results are technically corollaries to a theorem of Yarotsky, we expect that the paradigm of systems with a locally gapped ground state that is completely insensitive to the form of the Hamiltonian elsewhere extends to other situations and has important physical consequences.
\end{abstract}

\section{Introduction}	\label{sec:intro}

We consider weakly interacting quantum spin systems on finite subsets~\(\Lambda \) of the lattice~\(\Z^\nu\)\!, \(\nu\in\N\), described by a self-adjoint Hamiltonian
\begin{equation} \label{eq:introhamiltonian}
    H= H_0  + H_\mathrm{int}  \,,
\end{equation}
which is composed of a non-interacting part~\(H_0\) and an interacting part~\(H_\mathrm{int}\).
The non-interacting Hamiltonian~\(H_0\) is a sum of non-negative on-site Hamiltonians~\(h_x\), \(x\in \Lambda\).
Each~\(h_x\) is  assumed to have a non-degenerate ground state with ground state energy~\(0\) and spectral gap of size at least~\(g\) above the ground state.
The interaction Hamiltonian~\(H_\mathrm{int}\) is a sum of interaction terms~\(\Phi_x\) of finite range~\(R\) and of small uniformly bounded norm \(\norm{\Phi_x}\).
We show that for such Hamiltonians a strong version of the \emph{local perturbations perturb locally} (LPPL) principle holds:
For any self-adjoint perturbation~\(P\)\!, supported in a region \(X \subset \Lambda\), any ground state~\(\rho_P\) of the perturbed Hamiltonian \(H + P\) agrees with the ground state~\(\rho\) of the unperturbed Hamiltonian~\(H\) when tested against observables~\(A\) supported in a region \(Y\subset \Lambda\) up to an error that is exponentially small in the distance~\(\dist(Y,X)\).
More precisely, Theorem~\ref{thm:LPPL} states that there are positive constants \(c, c_1, c_2>0 \) depending only on~\(R\) and~\(g\), but not on~\(\Lambda\), \(A\), \(H\) or~\(P\)\!, such that whenever \(\norm{\Phi_x} \leq c\) for all \(x\in \Lambda\), it holds that
\begin{equation} \label{eq:introbound}
    \abs[\big]{\trace[\big]{\paren{ \rho_P  -  \rho  } A}}
    \, \le \,
    \E^{c_1 \abs{Y}} \, \norm{A} \, \E^{-c_2   \dist(Y,X) } \,.
\end{equation}
Note that the uniformity of the error estimate with respect to the system size~\(\abs{\Lambda}\) is one key aspect which makes this estimate non-trivial.
Note also, that the bound on~\(\norm{\Phi_x}\) implies that~\(H\) has a gap above its unique ground state~\(\rho\) as we discuss below.
However, for our result we neither require nor actually have any uniform lower bound on the gap above the possibly non-unique ground state~\(\rho_P\) of the perturbed Hamiltonian \(H + P\)\!.

As a corollary of our main theorem,  we show that a bound of the form~\eqref{eq:introbound} also holds for systems that have the appropriate structure of gapped on-site terms and weak interactions only locally in some region of space.
In particular, this shows that the notion of a locally gapped ground state, which is completely insensitive to the form of the Hamiltonian elsewhere, is perfectly valid in this setup.

The LPPL-principle was coined by Bachmann, Michalakis, Nachtergaele, and Sims in~\cite{bachmann2012automorphic}, where a similar estimate with subexponential decay was proven.
While their result covers much more general interacting quantum spin systems, it requires the gap above the ground state to remain open also for the perturbed Hamiltonian \(H + P\)\!.
More precisely, it relies on connecting \(H(0) := H\) with \(H(1) := H + P\) by a continuous path \([0,1] \ni t \mapsto H (t)\) in the space of Hamiltonians, such that the gap above the ground state of \(H (t)\) remains open uniformly along the whole path.
Then the locality of the quasi-adiabatic evolution introduced by Hastings and Wen in~\cite{hastings2005quasiadiabatic} can be used to prove the result.
Their subexponential bound was improved to exponential precision for finite-range interactions by De~Roeck and \LTskip{Sch\"utz} in~\cite{deRoeckLPPL}.
See also~\cite{nachtergaele2019quasi,nachtergaele2020quasi} for recent developments.

While we prove the strong version of the LPPL-principle only for weakly interacting spin systems, we expect it to hold somewhat more generally.
For example, we expect it to hold for fermions on the lattice with weak finite range interactions, a physical setup where the strong LPPL-principle would have important consequences.
It would imply that a gapped ground state for such a system with periodic boundary conditions remains unchanged in the bulk when introducing open boundary conditions that may close the global gap due to the emergence of edge states.
And as a consequence, it would also explain why the adiabatic response to external fields in the bulk of such systems is not affected by edge states that close the gap,
see~\cite{bachmann2018adiabatic, monaco2017adiabatic, teufel2020non, henheikteufel20202, henheikteufel20203} for related results.
However, it is known that the strong LPPL-principle cannot hold in general, but requires further conditions on the unperturbed ground state sector such as local topological quantum order (LTQO)~\cite{MZ13,nachtergaele2021stability}.

Shortly before resubmitting the final version of this article, Bachmann et~al.\ published a preprint containing a closely related result.
In~\cite{bachmann2021} they prove an LPPL-bound as in~\eqref{eq:introbound}, but with subexponential decay, assuming LTQO for a unique frustration-free gapped ground state of the unperturbed Hamiltonian which has no long-range entanglement.

Our result is a corollary of a result by Yarotsky~\cite{yarotsky2005uniqueness} (see Theorem~\ref{thm:Yarotsky} below), which provides a bound on the difference of so-called finite volume ground states in quantum spin systems described by Hamiltonians of the form~\eqref{eq:introhamiltonian}.
His aim and main result in that work was to show the uniqueness of the ground state of such systems in the thermodynamic limit.
In a different work Yarotsky~\cite{yarotsky2004perturbations} has shown that Hamiltonians of the form~\eqref{eq:introhamiltonian} with \(\norm{\Phi}<c\) indeed have a unique ground state separated by a gap \(\tilde g > 0\) from the rest of the spectrum, with \(\tilde g\) independent of \(\Lambda\) (see~\cite{DS,FP, hastings2019stability} for similar results).
Closely related to the stability of the gap is  the stability of phase diagrams at low temperatures, see~\cite{borgs, datta1, datta2}.

\paragraph{Acknowledgement}
J.\,H.~acknowledges partial financial support by the ERC Advanced Grant ``\LTskip{RMTBeyond}” No.\,101020331.
S.\,T.~thanks \LTskip{Marius Lemm} and \LTskip{Simone Warzel} for very helpful comments and discussions and \LTskip{J\"urg Fr\"ohlich} for references to the literature.

\section{Main results}\label{sec:result}

Consider the lattice \(\mathbb{Z}^\nu\) for fixed \(\nu \in \mathbb{N}\) equipped with the \(\ell^1\)-metric \(d \colon \Z^\nu\times\Z^\nu \to \mathbb{N}_0\) and define
\(\mathcal{P}_0(\Z^\nu) = \Set{ \Lambda \subset \Z^\nu \given \abs{\Lambda} < \infty }\), where~\(\abs{\Lambda}\) denotes the cardinality of~\(\Lambda\).
With each site \(x \in \Z^\nu\) one associates a (possibly infinite dimensional) Hilbert space~\(\HSpace_x\).
For \(\Lambda \in \mathcal{P}_0(\Z^\nu)\) set \(\HSpace_{\Lambda}=\bigotimes_{x \in \Lambda} \HSpace_x\) and denote the algebra of bounded linear operators on \(\HSpace_{\Lambda}\) by \(\Alg_{\Lambda}=\mathcal{B}(\HSpace_{\Lambda})\).
Due to the tensor product structure, we have \(\Alg_{\Lambda}=\bigotimes_{x \in \Lambda} \mathcal{B}(\HSpace_x)\).
Hence, for \(\Lambda'\subset\Lambda \in \mathcal{P}_0(\Z^\nu)\), any \(A \in \Alg_{\Lambda'}\) can be viewed as an element of \(\Alg_{\Lambda}\) by identifying \(A\) with \(A\otimes \unit_{\Lambda\setminus \Lambda'} \in \Alg_{\Lambda}\), where \(\unit_{\Lambda \setminus \Lambda'}\) denotes the identity in \(\Alg_{\Lambda \setminus \Lambda'}\).
Note that
\begin{equation*} \label{eq:commutator}
    \commutator{A,B} = 0
    \quad \text{for all} \quad A \in \Alg_\Lambda \,, \ B \in \Alg_{\Lambda'}
    \quad \text{with} \quad \Lambda \cap \Lambda' = \emptyset\,.
\end{equation*}

Similarly, we will also denote by \(K\) the closure of \(\unit_{\Lambda \setminus \Lambda'} \otimes K\) on \(\HSpace_{\Lambda \setminus \Lambda'} \otimes D(K)\) for any self-adjoint operator \(K\) on \(\HSpace_{\Lambda'}\).
Here and in the following, \(D(K)\) denotes the domain of the operator \(K\).

Our main result will be formulated for a Hamiltonian
\begin{equation*}
    H = H_0 + H_\mathrm{int} \in \Alg_{\Lambda}
\end{equation*}
that is composed of a non-interacting part~\(H_0\) and an interacting part \(H_\mathrm{int}\).
The non-interacting part~\(H_0\) is assumed to be of the form
\begin{equation*}
    H_0 = \sum_{x \in \Lambda} h_x\,,
\end{equation*}
where each \(h_x\) is a non-negative self-adjoint (possibly unbounded) operator on \(\HSpace_x\) with a unique gapped ground state \(\psi_x \in D(h_x)\) satisfying
\begin{equation} \label{eq:gap}
    h_x \psi_x = 0
    \quad \text{and} \quad
    h_x\big\vert_{D(h_x) \ominus \psi_x}  \geq  g\,,
\end{equation}
for some fixed~\(g>0\).
The latter means that \(\braket{\varphi_x,(h_x-g\unit_x)\varphi_x} \geq 0\) for all \(\varphi_x \in D(h_x)\) with \(\braket{\psi_x,\varphi_x}=0\).
In other words, all Hamiltonians~\(h_x\) have a spectral gap of size at least~\(g\) above the bottom of their spectrum.
The interacting part is of the form
\begin{equation*}
    H_\mathrm{int}  =  \sum_{x \in \Lambda} \Phi_x\,,
\end{equation*}
with \(\Phi_x\in \Alg_{b_x(R)}\) self-adjoint for each \(x \in \Lambda\) and some fixed \(R\in\N\).
Here \(b_x(R) := {\Set{y \in \Lambda \given d(x,y) \leq R}}\) denotes the \(\ell^1\)-ball with radius~\(R\) centered at \(x \in \Lambda\).
We set
\begin{equation*}
    \tnorm{\Phi} := \sup_{x \in \Lambda} \, \norm{\Phi_x} \,.
\end{equation*}

\begin{defi}\label{def:WISS} \thmheadfont{Weakly interacting spin system}\\
 For any \(\Lambda \in \mathcal{P}_0(\Z^\nu) \) we call a Hamiltonian \(H = H_0 + H_\mathrm{int}\) on~\(\HSpace_\Lambda\) with~\(H_0\) and~\(H_\mathrm{int}\) satisfying the above conditions a \emph{weakly interacting spin system on \(\Lambda\)} with  on-site gap~\(g\), interaction range~\(R\) and interaction strength~\(\tnorm{\Phi}\).
\end{defi}

We use the following definition for ground states and briefly explain how it is connected to the standard definition in Appendix~\ref{sec:groundstate}.

\begin{defi}
    \label{def:groundstate}
    Let \(\Lambda \in \mathcal{P}_0(\Z^\nu)\) and \(K\) be a self-adjoint and bounded below  operator on \(\HSpace_\Lambda\).
    We say that \(\commutator{K,A}\) is a bounded operator \(B\in\Alg_\Lambda\), whenever \(A\) leaves \(D(K)\) invariant and \(\commutator{K,A}=B\) on \(D(K)\).
    
    A state \(\rho \in \Alg_{\Lambda}\), i.e.~a positive semi-definite bounded operator with trace equal to one, is called a ground state of \(K\), if
    \begin{equation*} \label{eq:def-groundstate-algebra} 
        \trace*{ A^* \commutator{K,A} \,\rho}  \,\geq\, 0 \quad \text{for all \(A \in \Alg_{\Lambda}\) such that \(\commutator{K,A}\) is bounded} .
    \end{equation*}
\end{defi}

Our first main result is the following.

\begin{thm} \label{thm:LPPL}  \thmheadfont{The strong LPPL-principle }\\
    Let \(R\in \N\) and \(g>0\).
    There exist constants \(c, c_1, c_2>0\), such that for any \(\Lambda\in \mathcal{P}_0(\Z^\nu)\) and any weakly interacting spin system \(H = H_0 + H_\mathrm{int}\) on~\(\Lambda\) with on-site gap at least~\(g\), interaction range~\(R\), and interaction strength~\(\tnorm{\Phi} \leq c\) the following holds:

    Let \(X\subset\Lambda\) be non-empty and \(P\) be a symmetric operator on \(\HSpace_X\) such that \(P\) is relatively bounded with respect to \(H\) with \(H\)-bound less than one.
    Set \(H_P = H + P\)\!.
    Then for any ground state~\(\rho\) of~\(H\), any ground state~\(\rho_P\) of~\(H_P\), and all \(A \in \Alg_Y\) with \(Y \subset \Lambda\) it holds that
    \begin{equation} \label{eq:thm}
        \abs[\big]{\trace[\big]{ \paren{\rho_P -\rho } A}}
        \,\le\,
        \E^{c_1 \abs{Y}} \, \norm{A} \, \E^{-c_2 \dist(Y,X)} \,.
    \end{equation}
\end{thm}

\medskip

Under the assumptions of the theorem, Yarotsky has proven in~\cite{yarotsky2004perturbations} that~\(H\) has a unique ground state \(\rho\), whenever \(c>0\) is small enough.\footnote{
Note that the systems for which Yarotsky proves existence and uniqueness of the ground state in~\cite{yarotsky2004perturbations} differ slightly from our definition of weakly interacting spin systems in the treatment of interaction terms near the boundary of the domain.
To obtain the same result with our definition, one extends the Hamiltonian \(H\) to \(\Omega\supset\Lambda\) as in the proof of Theorem~\ref{thm:LPPL}, applies the result from~\cite{yarotsky2004perturbations} and restricts the resulting ground state to \(\Lambda\) by taking the partial trace.
}
In the following we will assume that this is the case.

For~\(X\) at the edge of~\(\Lambda\), the perturbation~\(P\) can be employed to realize all kinds of boundary conditions, e.g.\ if \(\Lambda = \{ -M,\dotsc,M \}^\nu\) is a box, periodic boundary conditions can be modeled by some~\(P\) connecting opposite sites in~\(\Lambda\).
Therefore, if~\(X\) is at the edge, one can take the thermodynamic limit \({\Lambda \nearrow \Z^\nu}\) in~\eqref{eq:thm} and conclude that there exists a unique ground state~\(\rho\), i.e.\ a normalized positive functional, on the \(C^*\)-algebra of quasi-local observables
\(\Alg = \LTskip{\overline{\Aloc}^{\raisebox{-2pt}{\(\scriptstyle\norm{\cdot}\)}}}\),
independent of the imposed boundary conditions for the finite systems.
This uniqueness of ground states for the infinite system was the main result  of~\cite{yarotsky2005uniqueness} and has been shown by Yarotsky based on Theorem~\ref{thm:Yarotsky}, which we quote below.

As mentioned in the introduction, we expect a similar strong LPPL-principle to hold also for  fermionic lattice systems with weak finite range interactions.
As discussed in~\cite{henheikteufel20202, henheikteufel20203}, this would have important consequences for linear response and adiabatic theorems for systems with a gap only in the bulk.

Our second main result is a local version of Theorem~\ref{thm:LPPL}, where we assume the on-site gap and the weak interaction only locally.

\begin{defi} \label{def:WISS-in-region}\thmheadfont{Locally weakly interacting spin system}\\
    For any \(\Lambda \in \mathcal{P}_0(\Z^\nu)\) and \(\Lambda'\subset\Lambda\) we say that a self-adjoint operator \(H\) on \(\HSpace_\Lambda\) is \emph{weakly interacting in the region~\(\Lambda'\)} with on-site gap~\(g\), range~\(R\) and strength~\(s\), if and only if there exists a weakly interacting spin system \(\tilde H = \tilde H_0 + \tilde H_\mathrm{int}\) on \(\Lambda\) with on-site gap~\(g\), range~\(R\) and strength~\(\tnorm{\Phi}=s\) such that \( H - \tilde H = \unit_{\HSpace_{\Lambda'}} \otimes \, Q\) with \(Q\) a possibly unbounded symmetric operator on \(\HSpace_{\Lambda\setminus\Lambda'}\) such that \(Q\) is infinitesimally \(\tilde H\)-bounded.
\end{defi}

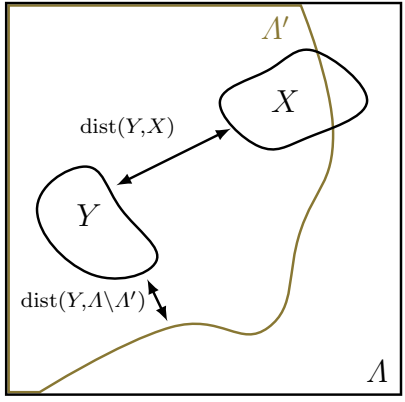
\begin{SCfigure}[1.3]
    \centering
    \begin{tikzpicture}[scale=.43, myLine]
        \newcommand{\pathouter}{(-6.2,6.2) rectangle (6.2,-6.2)}
        \newcommand{\pathL}{(-6,6) rectangle (6,-6)}
        \newcommand{\pathLp}{(-6,6) -- plot[smooth, tension=0.8] coordinates{(3,6) (4,2) (3,-1) (2,-4) (-1,-4) (-5,-6)} -- (-6,-6) -- cycle}
        \newcommand{\posX}{(2.5,3)}
        \newcommand{\pathX}{plot[smooth cycle, tension=.7]%
            coordinates{+(0:2.5) +(60:1.8) +(120:1.1) +(180:2) +(240:1.6) +(300:1.3)}}
        \newcommand{\posY}{(-3.5,-.5)}
        \newcommand{\pathY}{plot[smooth cycle, tension=0.8] coordinates{+(10:1) +(95:1.5) +(175:1.6) +(270:1.8) +(325:2.5)}}

        \begin{scope}[myLine=2, even odd rule, draw=myColor] 
            \clip \pathLp;
            \draw \pathLp;
        \end{scope}
        \node[anchor=north east, myColor] at (3,6) {\(\Lambda'\)}; 

        \draw \posX{} node {\(X\)} \pathX{}; 
        \draw \posY{} node {\(Y\)} \pathY{}; 

        \begin{scope}[myLine=2, even odd rule] 
            \clip \pathouter \pathL;
            \draw \pathL;
        \end{scope}
        \draw (6,-6) node[anchor=south east] {\(\Lambda\)};

        \draw[latex-latex, shorten <=0.4cm, shorten >=0.8cm] \posY{} ++(90:.5) -- node[anchor=south east] {\(\scriptstyle\dist(Y,X)\)} \posX{};
        \draw[latex-latex, shorten <=2pt, shorten >=3pt] \posY{} ++(-47:2.5) -- node[anchor=east] {\(\scriptstyle\dist(Y,\Lambda\setminus\Lambda')\)} (-1,-4);
    \end{tikzpicture}
    \caption{Depicted is the setting from Corollary~\ref{thm:LPPL-local-gaps}.
        The system~\(H\) defined on~\(\Lambda\) is assumed to be weakly interacting and to have an on-site gap in \(\Lambda'\subset\Lambda\).
        For any perturbation~\(P\) acting on \(X\subset\Lambda\), ground states of~\(H\) and \(H+P\) agree in regions~\(Y\) away from~\(X\) and \(\Lambda\setminus\Lambda'\).
    }
    \label{fig:cor}
\end{SCfigure}

\begin{cor} \label{thm:LPPL-local-gaps}  \thmheadfont{The strong LPPL-principle for local gaps}\\
    Let~\(R \in \N\), \(g>0\), and \(c, c_1, c_2>0\) be the constants from Theorem~\ref{thm:LPPL}.
    Then for any \(\Lambda \in \mathcal{P}_0(\Z^\nu)\), \(\Lambda'\subset\Lambda\), and any self-adjoint operator \(H\) on \(\HSpace_\Lambda\) which is weakly interacting in the region \(\Lambda'\) with on-site gap at least~\(g\), range~\(R\) and strength~\(s \leq c\) the following holds:

    Let \(X\subset\Lambda\) be non-empty and \(P\) be a symmetric operator on \(\HSpace_X\) such that \(P\) is relatively bounded with respect to \(H\) with \(H\)-bound less than one.
    Set \(H_P = H + P\) (see Figure~\ref{fig:cor}).
    Then for any ground state~\(\rho \) of~\(H \), any ground state~\(\rho _P\) of~\(H _P\), and all \(A \in \Alg_Y\) with \(Y \subset \Lambda'\) it holds that
    \begin{equation*}
        \abs[\big]{\trace[\big]{ \paren{\rho _P   -\rho } A}}
        \,\le\,
        2 \, \E^{c_1 \abs{Y}} \, \norm{A} \, \E^{-c_2 \min \List{ \dist(Y,X), \dist(Y,\Lambda\setminus \Lambda')}}  \,.
    \end{equation*}
\end{cor}

\begin{proof}
    Let~\(\tilde{H}\) and~\(Q\) be as in Definition~\ref{def:WISS-in-region}.
    Then~\(\tilde H\), \(\tilde H + Q\) and~\(\tilde H + Q + P\) are self-adjoint.
    For the latter this follows, because also \(Q+P\) is relatively bounded with respect to \(\tilde H\) with \(\tilde H\)-bound less than one.
    This is not obvious, but the proof is a straightforward calculation that we skip.
    
    Let~\(\tilde \rho\) be a ground state of~\(\tilde{H}\), see the comment after Theorem~\ref{thm:LPPL} for existence.
    Then the triangle inequality and two applications of Theorem~\ref{thm:LPPL} yield
    \begin{align*}
        \abs[\big]{\trace[\big]{ \paren{\rho _P   -\rho } A}}
        \, &  \le \,
        \abs[\big]{\trace[\big]{ \paren{\rho _P   -\tilde \rho } A}} +    \abs[\big]{\trace[\big]{ \paren{\rho    -\tilde \rho } A}}\\
        \, &  \le \,
        \E^{ c_1 \abs{Y}} \, \norm{A} \, \paren[\big]{ \E^{-c_2 \dist(Y,X \cup (\Lambda\setminus\Lambda')) } + \E^{-c_2 \dist(Y,\Lambda\setminus \Lambda') } }\,.\qedhere
    \end{align*}
\end{proof}

\section{Proof} \label{sec:proof}

The proof of Theorem~\ref{thm:LPPL} is essentially  a reinterpretation of  a result by Yarotsky~\cite{yarotsky2005uniqueness}.
Since we only deal with finite volumes, we modify Yarotsky's notion of \emph{finite volume ground states} to \emph{ground states in the bulk}.
To make the arguments as transparent as possible, we will add superscripts to Hamiltonians and states indicating on which subset of \(\Z^\nu\) they are defined.
These superscripts are also used to distinguish different operators and states.
From now on let
\begin{equation*}
    \domainH_\Lambda := \Set{A\in\Alg_\Lambda \given \commutator{H_0^\Lambda,A} \text{ is bounded}}
\end{equation*}
and note, that also \(\Set{A\in\Alg_\Lambda \given \commutator{H_0^\Lambda+K,A} \text{ is bounded\LTadd{ }}} = \domainH_\Lambda\) for all bounded operators \(K\in\Alg_\Lambda\).

\begin{defi} \label{def:groundstatebulk}\thmheadfont{Ground states in the bulk}\\
    Let \(R \in \mathbb{N}\), \(\Lambda_* \subset \Lambda \in \mathcal{P}_0(\Z^\nu)\) and \(H^{\Lambda_*} = H_0^{\Lambda_*} + H_\mathrm{int}^{\Lambda_*} \in \Alg_{\Lambda_*}\) be a weakly interacting spin system on \(\Lambda_*\) with range~\(R\).
    Then we call
    \begin{equation*}
        \Lambda_*^\circ := \Set{ x \in \Lambda_* \given \dist(x,\Z^\nu\setminus \Lambda_*) > 2R}
    \end{equation*}
    the \emph{bulk} of the Hamiltonian \(H^{\Lambda_*}\) and any state \(\rho^\Lambda \in \Alg_{\Lambda}\) satisfying
    \begin{equation*}
        \trace*{\rho^\Lambda \, A^* \commutator{H^{\Lambda_*},A} }
        \geq  0
        \quad \text{for all} \quad A \in \domainH_{\Lambda_*^\circ}
    \end{equation*}
    a  \emph{ground state in the bulk of \(H^{\Lambda_*}\)}.
\end{defi}

Our proof is based on the following theorem due to Yarotsky~\cite{yarotsky2005uniqueness}.

\begin{thm} \thmcitefont {(\cite[Theorem 2]{yarotsky2005uniqueness})} \label{thm:Yarotsky}
    Let~\(R\in \N\) and~\(g>0\).
    There exist constants \(c, c_1, c_2>0\) such that for any \(\Lambda_* \in \mathcal{P}_0(\Z^\nu)\), and any weakly interacting spin system \(H^{\Lambda_*} = H_0^{\Lambda_*} + H_\mathrm{int}^{\Lambda_*}\) on~\(\Lambda_*\) with on-site gap at least~\(g\), range~\(R\) and interaction strength~\(\tnorm{\Phi}\leq c\) the following holds:

    Let \(\Lambda \in \mathcal{P}_0(\Z^\nu)\) be such that \(\Lambda_* \subset \Lambda\).
    Then for any two ground states~\(\rho^\Lambda_1\) and~\(\rho^\Lambda_2 \in \Alg_\Lambda\) in the bulk of~\(H^{\Lambda_*}\) in the sense of Definition~\ref{def:groundstatebulk}, \(Y \subset \Lambda_*\), and~\(A \in \Alg_Y\) it holds that
    \begin{equation*}
        \abs[\big]{ \trace[\big]{\paren{ \rho^\Lambda_1  - \rho^\Lambda_2 } A}}
        \, \le \,
        \E^{c_1 \abs{Y}} \, \norm{A} \, \E^{-c_2\,  \dist(Y, \Z^\nu \setminus \Lambda_*^\circ)}  \,.
    \end{equation*}
\end{thm}

Note that the set denoted by~\(\Lambda\) in~\cite[Theorem~2]{yarotsky2005uniqueness} corresponds to our set~\(\Lambda_*\).
Note, moreover, that any ground state~\(\rho^\Lambda\) in the bulk of~\(H^{\Lambda_*}\) trivially defines a finite-volume ground state \(A \mapsto \trace{\rho^\Lambda \, (A\otimes \unit_{\Lambda\setminus\Lambda_*})}\) of~\(H^{\Lambda_*}\) in the sense of~\cite[Definition~2]{yarotsky2005uniqueness}.
Allowing an arbitrary on-site gap~\(g>0\) instead of~\(g=1\), as in~\cite{yarotsky2005uniqueness}, is achieved by simple scaling.

\begin{lem} \label{lem:ground-state-to-ground-state-in-bulk}
    Let \(R \in \mathbb{N}\), \(\Lambda_*\subset \Lambda \in  \mathcal{P}_0(\Z^\nu)\) and \(H^{\Lambda} = H_0^{\Lambda} + H_\mathrm{int}^{\Lambda} \in \Alg_{\Lambda}\) be a weakly interacting spin system.
    Then the canonical restriction of \(H^\Lambda\) to \(\Lambda_*\) defined by
    \begin{equation*}
        H^{\Lambda}|_{\Lambda_*}
        = H_0^{\Lambda}|_{\Lambda_*} + H_\mathrm{int}^{\Lambda}|_{\Lambda_*}
        := \sum_{x \in \Lambda_*} h_x + \sumstack[lr]{x \in \Lambda_*:\\ \dist(x,\Lambda\setminus\Lambda_*)>R} \Phi_x
    \end{equation*}
    is a weakly interacting spin system on \(\Lambda_*\) with the same on-site gap, range and strength and has the following property:
    For any symmetric operator \(Q\) on \(\HSpace_{\Lambda \setminus \Lambda^\circ_*}\) such that \(Q\) is relatively bounded with respect to \(H^\Lambda\) with \(H^\Lambda\)-bound less than one, any ground state of \(H^\Lambda + Q\) is also a ground state in the bulk of~\(H^{\Lambda}|_{\Lambda_*}\).
\end{lem}

\begin{proof}
    It is clear that~\(H^{\Lambda}|_{\Lambda_*}\) is a weakly interacting spin system on~\(\Lambda_*\).
    A simple calculation shows, that \(Q\) is also relatively bounded with respect to \(H_0^{\Lambda \setminus \Lambda^\circ_*}=\sum_{x \in \Lambda \setminus \Lambda^\circ_*} h_x\) with \(H_0^{\Lambda \setminus \Lambda^\circ_*}\)-bound less than one.
    Hence, \(K:= (H^\Lambda - H^{\Lambda}|_{\Lambda_*} + Q)\) is a self-adjoint operator on~\(\HSpace_{\Lambda\setminus\Lambda^\circ_*}\).
    Moreover, any~\(A\in\Alg_{\Lambda_*^\circ}\) leaves invariant the domain of~\(\unit_{\Lambda_*^\circ} \otimes K\) and satisfies
    \begin{equation*}
        \commutator*{ \unit_{\Lambda_*^\circ} \otimes K, A \otimes \unit_{\Lambda\setminus\Lambda_*^\circ}} = 0.
    \end{equation*}
    
    Similarly, for all~\(A\in\domainH_{\Lambda_*^\circ}\), \(\commutator{H^\Lambda + Q, A}\) is bounded and satisfies
    \begin{equation*}
        \commutator{H^\Lambda + Q, A} = \commutator{H^{\Lambda}|_{\Lambda_*}, A}.
    \end{equation*}
    Therefore, any ground state of \(H^\Lambda+Q\) is also a ground state in the bulk of~\(H^{\Lambda}|_{\Lambda_*}\).
\end{proof}

\begin{SCfigure}[1.3]
    \centering
    \begin{tikzpicture}[scale=.43, myLine]
        \newcommand{\pathL}{(-6,6) rectangle (6,-6)}
        \newcommand{\pathLo}{(-5,5) rectangle (5,-5)}
        \newcommand{\pathLpo}{(\((-5,5) + (\offset,-\offset)\)) rectangle (\((5,-5) + (-\offset,\offset)\))}
        \newcommand{\posX}{(-2.4,-2.5)}
        \newcommand{\pathX}[1]{plot[smooth cycle, tension=.8]%
            coordinates{+(0:#1+1.5) +(60:#1+1) +(120:#1+1.2) +(180:#1+.8) +(240:#1+1) +(300:#1+1.3)}}
        \newcommand{\pathY}{plot[smooth cycle, tension=0.8] coordinates{+(10:.6) +(95:2) +(175:1.6) +(270:1.8) +(325:2.5)}}

        \newcommand{\drawXYL}[1]{
            \draw \posX{} node {\(X\)} \pathX{0};
            \draw (2,1.5) +(225:.6) node {\(Y\)} \pathY;
            \draw \pathL node[anchor=south east] {#1};
        }

        \begin{scope}
            \begin{scope}[myColor]
                \begin{scope}[even odd rule, myLine=2, fill=myColor!10, draw=myColor]
                    \clip \pathLo (0,0) [shift={\posX{}}] \pathX{1};
                    \path[draw, fill] \pathLo (0,0) [shift={\posX{}}] \pathX{1};
                    \node[anchor=south, inner sep=2pt] at (0,-5) {\(\Lambda_*^\circ\)};
                \end{scope}
            \end{scope}
            \begin{scope}[myLine=2, even odd rule]
                \clip \pathL \pathLo;
                \draw \pathLo;
            \end{scope}
            \node[anchor=south east] at (5,-5) {\(\Lambda^\circ\)};
        \end{scope}
        \drawXYL{\(\Lambda\)}
    \end{tikzpicture}
    \caption{Depicted is the setting from the proof of Proposition~\ref{thm:LPPL-intermediate}.
        The subset~\(X\subset\Lambda\) is the region where the perturbation~\(P\) acts, and we choose \({\Lambda_*=\Lambda \setminus X}\).
        The shaded region~\(\Lambda_*^\circ\) is the bulk of~\(H^{\Lambda_*}\)\!.
        \({Y\subset\Lambda_*^\circ}\) is the support of the observable~\(A\).
        This indicates why~\eqref{eq:distance} holds.
    }
    \label{fig:prop}
\end{SCfigure}
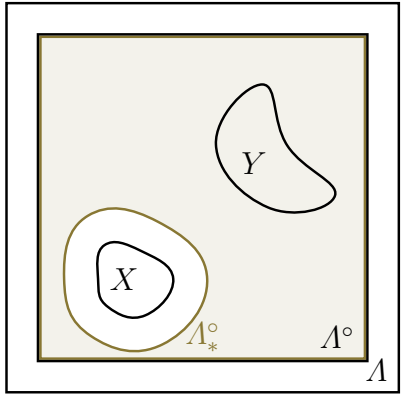

Before we prove Theorem~\ref{thm:LPPL}, let us give an intermediate result, which follows rather directly from
Theorem~\ref{thm:Yarotsky} and Lemma~\ref{lem:ground-state-to-ground-state-in-bulk}.

\begin{prop} \label{thm:LPPL-intermediate}
    Let~\(R\in \N\) and~\(g>0\).
    There exist constants \(c, c_1, c_2>0\) such that for any \(\Lambda\in \mathcal{P}_0(\Z^\nu)\) and any weakly interacting spin system \(H^\Lambda = H^\Lambda_0 + H^\Lambda_\mathrm{int}\) on \(\Lambda\) with on-site gap at least~\(g\), interaction range~\(R\), and interaction strength~\(\tnorm{\Phi} \leq c\) the following holds:
    
    Let \(X\subset\Lambda\) be non-empty and \(P\) be a symmetric operator on \(\HSpace_X\) such that \(P\) is relatively bounded with respect to \(H^\Lambda\) with \(H^\Lambda\)-bound less than one.
    Set \(H^\Lambda_P = H^\Lambda + P\)\!.
    Then for any ground state~\(\rho^\Lambda\) of~\(H^\Lambda\), any ground state~\(\rho^\Lambda_P\) of~\(H^\Lambda_P\), and all \(A \in \Alg_Y\) with \(Y \subset \Lambda\) it holds that
    \begin{equation*}
        \abs[\big]{\trace[\big]{ \paren{\rho^\Lambda_P   - \rho^\Lambda } A}}
        \, \le \,
        \E^{c_1 \abs{Y}} \, \norm{A} \, \E^{-c_2 \min\List*{ \dist(Y, \Z^\nu \setminus \Lambda^\circ), \, \dist(Y,X) - 2R}}  \,.
    \end{equation*}
\end{prop}

\begin{proof}
    Assume w.l.o.g.\ that \(Y \subset \Lambda^\circ\).
    Otherwise, the statement in Proposition~\ref{thm:LPPL-intermediate} is trivially satisfied after a possible adjustment of \(c_1\).

    Let \(\Lambda_* = \Lambda \setminus X\), and let \(H^{\Lambda}|_{\Lambda_*} \) be the canonical restriction of~\(H^\Lambda\) to~\(\Lambda_*\) as defined in Lemma~\ref{lem:ground-state-to-ground-state-in-bulk}.
    Then \(\Lambda_*^\circ \cap X = \emptyset\).
    We can assume   w.l.o.g.\ that \(\dist(X,Y) > 2R\) since otherwise the statement in Proposition~\ref{thm:LPPL-intermediate} is trivially satisfied after a possible adjustment of~\(c_1\).
    Then also \(Y \subset \Lambda_*^\circ\) (compare Figure~\ref{fig:prop}).
    By application of Lemma~\ref{lem:ground-state-to-ground-state-in-bulk} with \(Q = P\) and \(Q = 0\) we find that both, \(\rho_P^\Lambda\) and \(\rho^\Lambda\), are ground states in the bulk of \(H^{\Lambda}|_{\Lambda_*}\).
    Hence, Theorem~\ref{thm:Yarotsky} implies that
    \begin{equation*} \label{eq:proof}
        \abs[\big]{ \trace[\big]{\paren{\rho^\Lambda_P - \rho^\Lambda} A}} \, \le \, \E^{c_1 \abs{Y}} \, \norm{A} \, \E^{-c_2 \dist(Y, \Z^\nu \setminus \Lambda_*^\circ)} \,.
    \end{equation*}
    From
    \begin{equation*}
        \Z^\nu \setminus \Lambda_*^\circ = (\Z^\nu \setminus \Lambda^\circ) \cup \Set{ x \in \Z^\nu \given \dist(x,X) \le 2R }
    \end{equation*}
    we immediately conclude that
    \begin{equation}\label{eq:distance}
        \dist(Y, \Z^\nu \setminus \Lambda_*^\circ) = \min \List*{  \dist(Y, \Z^\nu \setminus \Lambda^\circ), \, \dist(Y, X) - 2R }\,,
    \end{equation}
    which yields the claim.
\end{proof}

We now extend this result to obtain Theorem~\ref{thm:LPPL}.

\begin{proof}[Proof of Theorem~\ref{thm:LPPL}]
    In the following, we add superscripts~\(\Lambda\) to the Hamiltonians and states from the statement of Theorem~\ref{thm:LPPL}.

    Let \(\Omega \in \mathcal{P}_0(\Z^\nu)\) be such that \(\Lambda \subset \Omega\).
    For each \(x \in \Omega \setminus \Lambda\) let \(h_x \in \Alg_{\List{x}}\) be a self-adjoint operator with gap at least \(g\) and non-degenerate ground state \(\psi_x\) satisfying~\eqref{eq:gap}.
    Then \(\rho^{\Omega \setminus \Lambda} = \bigotimes_{x\in\Omega \setminus \Lambda} \ket{\psi_x}\bra{\psi_x}\) is the ground state of
    \begin{equation*}
        H_0^{\Omega\setminus\Lambda} := \sumstack[lr]{x\in\Omega\setminus\Lambda} h_x\,.
    \end{equation*}
    Moreover, \(\rho^{\Omega}:=\rho^{\Lambda} \otimes \rho^{\Omega \setminus \Lambda}\) is a
    ground state of \(H^\Omega := H^\Lambda + H_0^{\Omega\setminus\Lambda}\) which is a weakly interacting spin system on \(\Omega\) with on-site gap at least~\(g\), range~\(R\), and interaction strength~\(\tnorm{\Phi}\).
    And also \(\rho_P^{\Omega}:=\rho_P^{\Lambda} \otimes \rho^{\Omega \setminus \Lambda}\) is a ground state of \(H_P^\Omega := H_P^\Lambda + H_0^{\Omega\setminus\Lambda} = H^\Omega + P\)\!.

    According to Proposition~\ref{thm:LPPL-intermediate} we have
    \begin{equation*}
        \abs[\big]{\trace[\big]{ \paren{\rho^\Omega_P   - \rho^\Omega } A}}
        \, \le \,
        \E^{c_1 \abs{Y}} \, \norm{A} \, \E^{-c_2 \min\List*{ \dist(Y, \Z^\nu \setminus \Omega^\circ), \, \dist(Y,X) - 2R}}
    \end{equation*}
    for all \(A\in\Alg_Y\) and \(Y\subset \Omega\).
    By requiring \(Y\subset \Lambda\) we obtain
    \begin{equation*}
        \abs[\big]{\trace[\big]{ \paren{\rho^\Lambda_P   - \rho^\Lambda } A}}
        = \abs[\big]{\trace[\big]{ \paren{\rho^\Omega_P   - \rho^\Omega } A}}
        \le
        \E^{c_1 \abs{Y}} \, \norm{A} \, \E^{-c_2 \min\List*{ \dist(\Lambda, \Z^\nu \setminus \Omega^\circ), \, \dist(Y,X) - 2R}}.
    \end{equation*}
    Since this bound is independent of~\(\Omega\), we can choose~\(\Omega\)   sufficiently large such that  \( \dist(\Lambda, \Z^\nu \setminus \Omega^\circ) >  \dist(Y,X) - 2R\).
    Absorbing \(\E^{2c_2R}\) in~\(c_1\) yields the claim.
\end{proof}

\appendix

\section{Characterization of ground states}
\label{sec:groundstate}

In the following lemma we show that every ground state in the usual sense, i.e.\ every minimizer of the energy functional, is also a ground state according to Definition~\ref{def:groundstate}.
While Definition~\ref{def:groundstate} is often used as a characterization of ground states in the context of extended quantum lattice systems, we could not find any reference in the literature, which covers the statement of the following lemma also for unbounded operators.

\begin{lem}
    \label{lem:groundstate}
    Let \(\Lambda \in \mathcal{P}_0(\Z^\nu)\) and~\(K\) be a self-adjoint and bounded below  operator on \(\HSpace_\Lambda\).
    A state \(\rho \in \Alg_{\Lambda}\) is a ground state in the usual sense, i.e. 
    \begin{equation} \label{eq:lem-groundstate} 
         \trace{ K\, \rho}  \,\leq\, \trace{K\,\tilde \rho   }
        \quad \text{for all states \(\tilde \rho \in \Alg_{\Lambda}\)}\,,
    \end{equation}
    if and only if \(\range(\rho) \subset D(K)\) and
    \begin{equation} \label{eq:lem-groundstate-algebra} 
        \trace*{ A^* \commutator{K,A} \,\rho}  \,\geq\, 0 \quad \text{for all \(A \in \Alg_{\Lambda}\) such that \(\commutator{K,A}\) is bounded}.
    \end{equation}
    Here we adopt the convention that the trace of an operator that is not trace class is \(+\infty\).
\end{lem}

For bounded~\(K\) this implies that for any state~\(\rho\in\Alg_\Lambda\) the conditions~\eqref{eq:lem-groundstate} and~\eqref{eq:lem-groundstate-algebra} are equivalent.
And for unbounded~\(K\), any ground state in the usual sense is a ground state according to our Definition~\ref{def:groundstate}.
It could be that equivalence extends to unbounded operators, i.e.\ that~\eqref{eq:lem-groundstate-algebra}  implies~\(\range(\rho) \subset D(K)\).
However, we could not find a proof for this.

\begin{proof}[Proof of Lemma~\ref{lem:groundstate}]
Let \(E_0:= \inf\sigma(K)\) and let~\((\phi_n)\) be a \LTskip{Weyl} sequence for~\(E_0\), i.e.\ \(\phi_n \in D(K)\), \(\norm{\phi_n}=1\) and \(\norm{(K-E_0)\,\phi_n } \leq 1/n\) for all~\(n\in\N\). 

Assume that~\(\rho\) satisfies~\eqref{eq:lem-groundstate}.
Since \(\trace{K \, \ket{\phi_n}\bra{\phi_n}}  \leq E_0 + 1/n\), it follows that \(\trace{K\rho} = E_0\).
Hence, \(E_0\) is an eigenvalue of~\(K\) and the range of~\(\rho\) is contained in the ground state eigenspace.  
Let \(A\in \Alg_\Lambda\) such that \(\commutator{K,A}\) is bounded. Then   the operator \(A^*(K-E_0)A\) is non-negative and
\begin{equation*}    
    \trace[\big]{ A^*\commutator{K,A}\,\rho}
    = \trace[\big]{\rho^{\frac12} A^* \commutator{K,A} \, \rho^{\frac12}}
    = \trace[\big]{\rho^{\frac12} A^* (K-E_0) A \, \rho^{\frac12}}
    \geq 0
\end{equation*}
follows.

Now assume that~\(\rho\) is a ground state in the sense of Definition~\ref{def:groundstate}, denote by \(\rho = \sum_{i} \lambda_i\,\ket{\psi_i}\bra{\psi_i}\) a spectral decomposition of~\(\rho\) with~\(\psi_i\) normalized.
Since~\(\range(\rho) \subset D(K)\), also~\(\psi_i\in D(K)\).
The operator \(A_{n,j} := \ket{\phi_n} \bra{\psi_j}\) then has a bounded commutator with~\(K\) and inequality~\eqref{eq:lem-groundstate-algebra} yields 
\begin{equation*}
    0
    \leq \trace[\big]{ A_{n,j}^*\commutator{K,A_{n,j}}\,\rho}
    = \lambda_j \, \braket[\big]{ \phi_n, [K,A_{n,j}] \, \psi_j}
    \leq \lambda_j \, \paren[\big]{ E_0 + \tfrac1n - \braket{\psi_j,K\psi_j} }.
\end{equation*}
Thus, \(\braket{\psi_j,K\psi_j}\leq E_0\) for all~\(j\). Hence, \(\trace{K\rho} = E_0\) and~\(\rho\) is indeed a ground state of~\(K\).
\end{proof}



\begin{thebibliography}{26}


\bibitem{bachmann2012automorphic}
S.~Bachmann, S.~Michalakis, B.~Nachtergaele, and R.~Sims.
\newblock Automorphic equivalence within gapped phases of quantum lattice
systems.
\newblock \emph{Communications in Mathematical Physics} 309:835--871, 2012.



\bibitem{bachmann2018adiabatic}
S.~Bachmann, W.~De~Roeck, and M.~Fraas.
\newblock The adiabatic theorem and linear response theory for extended quantum
systems.
\newblock \emph{Communications in Mathematical Physics} 361:997--1027, 2018.



\bibitem{bachmann2021}
S.~Bachmann, W.~De~Roeck, B.~Donvil, and M.~Fraas.
\newblock Stability against large perturbations of invertible, frustration-free ground states.
Preprint available at \href{https://arxiv.org/abs/2110.11194}{\texttt{arXiv:2110.11194}} (2021).



\bibitem{borgs}
C.~Borgs, R.~Kotecky, and D.~Ueltschi.
\newblock Low temperature phase diagrams for quantum perturbations of classical spin systems.
\newblock \emph{ Communications in  Mathematical Physics} 181:409--446, 1996.


\bibitem{datta1}
N.~Datta, R.~Fernandez, and J.~Fr\"ohlich.
\newblock Low-temperature phase diagrams of quantum lattice systems. I\@.
Stability for quantum perturbations of classical systems with finitely-many ground states.
\newblock \emph{ Journal of Statistical Physics} 84:455–534, 1996.

\bibitem{datta2}
N.~Datta, R.~Fernandez, J.~Fr\"ohlich, and L.~Rey-Bellet.
\newblock Low-temperature phase diagrams of quantum lattice systems. II\@.
Convergent perturbation expansions and stability in systems with infinite degeneracy.
\newblock \emph{Helvetica Physica Acta} 69:752--820, 1996.


\bibitem{DS}
W.~De~Roeck and M.~Salmhofer.
\newblock  Persistence of exponential decay and spectral gaps for interacting fermions.
\newblock \emph{Communications in Mathematical Physics} 365:773--796, 2019.


\bibitem{deRoeckLPPL}
W.~De~Roeck and M.~Sch\"utz.
\newblock Local perturbations perturb---exponentially---locally.
\newblock \emph{Journal of Mathematical Physics} 56:061901, 2015.


\bibitem{FP}
J.~Fr\"ohlich and A.~Pizzo.
\newblock Lie-Schwinger block-diagonalization and gapped quantum chains.
\newblock \emph{Communications in Mathematical Physics} 375:2039--2069, 2020.


\bibitem{hastings2019stability}
M.~B.~Hastings.
\newblock The stability of free Fermi Hamiltonians.
\newblock \emph{Journal of Mathematical Physics} 60:042201, 2019.


\bibitem{hastings2005quasiadiabatic}
M.~B.~Hastings and X.-G.~Wen.
\newblock Quasiadiabatic continuation of quantum states: The stability of
topological ground-state degeneracy and emergent gauge invariance.
\newblock \emph{Physical Review B} 72:045141, 2005.


\bibitem{henheikteufel20202}
J.~Henheik and S.~Teufel.
\newblock Adiabatic theorem in the thermodynamic limit: Systems with a uniform gap.
\newblock \emph{Journal of Mathematical Physics} 63:011901, 2022.


\bibitem{henheikteufel20203}
J.~Henheik and S.~Teufel.
\newblock Adiabatic theorem in the thermodynamic limit: Systems with a gap in the bulk.
\newblock \emph{Forum of Mathematics, Sigma} 10:E4, 2022.


\bibitem{nachtergaele2019quasi}
B.~Nachtergaele, R.~Sims, and A.~Young.
\newblock Quasi-locality bounds for quantum lattice systems. I\@.
Lieb-Robinson bounds, quasi-local maps, and spectral flow automorphisms.
\newblock \emph{Journal of Mathematical Physics} 60:061101, 2019.


\bibitem{nachtergaele2020quasi}
B.~Nachtergaele, R.~Sims, and A.~Young.
\newblock Quasi-locality bounds for quantum lattice systems.
Part II\@.
Perturbations of frustration-free spin models with gapped ground states.
\newblock \emph{Annales Henri Poincar\'e}, Online First 2021.


\bibitem{nachtergaele2021stability}
B.~Nachtergaele, R.~Sims, and A.~Young.
\newblock Stability of the bulk gap for frus\-tra\-tion-free topologically ordered quantum lattice systems. \newblock Preprint available at \href{http://arxiv.org/abs/2102.07209}{\texttt{arXiv:2102.07209}} (2021).


\bibitem{MZ13}
S.~Michalakis and J.~Zwolak.
\newblock Stability of frustration-free Hamiltonians.
\newblock \emph{Communications in Mathematical Physics} 322:277--302, 2013.


\bibitem{monaco2017adiabatic}
D.~Monaco and S.~Teufel.
\newblock Adiabatic currents for interacting fermions on a lattice.
\newblock \emph{Reviews in Mathematical Physics} 31:1950009, 2019.


\bibitem{tasaki2020}
H.~Tasaki.
\newblock \emph{Physics and mathematics of quantum many-body systems.}
\newblock Springer, Singapore, 2020.


\bibitem{teufel2020non}
S.~Teufel.
\newblock Non-equilibrium almost-stationary states and linear response for  gapped quantum systems.
\newblock \emph{Communications in Mathematical Physics} 373:621--653, 2020.


\bibitem{yarotsky2004perturbations}
D.~Yarotsky.
\newblock Perturbations of ground states in weakly interacting quantum spin systems.
\newblock \emph{  Journal of Mathematical Physics} 45.6:2134--2152, 2004.


\bibitem{yarotsky2005uniqueness}
D.~Yarotsky.
\newblock Uniqueness of the ground state in weak perturbations of non-interacting gapped quantum lattice systems.
\newblock \emph{Journal of Statistical Physics} 118:119--144, 2005.

\enlargethispage{\baselineskip}

\end{thebibliography}
\end{document}